\documentclass[prd,preprint,showpacs,nofootinbib,tightenlines,groupedaddress]{revtex4}
\usepackage{epsfig}
\usepackage{amssymb}
\usepackage{amsfonts}
\usepackage{subfigure}
\usepackage{float}
\usepackage{latexsym,hyperref}
\graphicspath{{./Figures/}} 

\begin{document}
\preprint{HU-EP-06/07}

\title{Vortex free energy and deconfinement in center-blind discretizations 
of Yang-Mills theories}

\author{G.~Burgio\footnote{Address from September 1$^{\rm st}$ 2006: 
Universit\"at T\"ubingen, Institut f\"ur Theoretische Physik.}}

\author{M.~Fuhrmann}

\author{W.~Kerler}

\author{M.~M\"uller-Preussker}

\affiliation{Humboldt-Universit\"at zu Berlin, Institut f\"ur Physik, Berlin, 
Germany}

\date{\today}%

\begin{abstract}
Maximal 't~Hooft loops are studied in $SO(3)$ lattice gauge theory 
at finite temperature $T$. Tunneling barriers among twist sectors 
causing loss of ergodicity for local update algorithms are overcome 
through parallel tempering, enabling us to measure the vortex free 
energy $F$ and to identify a deconfinement transition at some 
$\beta_A^{crit}$. The behavior of $F$ below $\beta_A^{crit}$ shows 
however striking differences with what is expected from discretizations 
in the fundamental representation. 
\end{abstract}

\maketitle

Topology plays an important r\^ole in the non-perturbative dynamics of 
Yang-Mills theories. In particular the vacuum 
condensation of topological excitations might explain quark confinement and 
the existence of a mass gap through various scenarios  
described in the literature 
\cite{'tHooft:1974qc,Polyakov:1977fu,'tHooft:1979uj}. 
Some of these models allow to define
a topological order parameter for the finite $T$ deconfinement transition; 
for 't~Hooft magnetic vortices, classified by the 
first homotopy class $\mathbb{Z}_N$ 
of the continuum Yang-Mills gauge group $SU(N)/\mathbb{Z}_N$ 
\cite{'tHooft:1979uj}, the change in free 
energy $F = \Delta U -T \Delta S$ for their creation might play such r\^ole 
and has received broad attention, in particular in lattice discretizations
at zero and finite $T$ 
\cite{Kovacs:2000sy,deForcrand:2000fi,deForcrand:2001nd}. 
The main problem in such non-perturbative regularizations 
is that creating a vortex is equivalent to the introduction of a non-trivial
twist. For discretizations in the fundamental representation,
transforming under the {\it enlarged} gauge group $SU(N)$ ($N=2$ in this 
paper), this cannot be implemented dynamically but only via a modification 
of their boundary conditions (b.c.), the generalized partition function 
$\tilde{Z}$ being 
defined through the weighted sum of partition functions with fixed 
twisted b.c. Since each of them must be determined by 
independent simulations their relative weights can only be calculated through 
indirect means \cite{deForcrand:2000fi,deForcrand:2001nd}. 

Universality arguments are often cited to claim that results in lattice 
Yang-Mills theories will not depend on the discretization chosen. A 
natural alternative in calculating $F$ would therefore be to directly 
discretize the theory in the $SU(N)/{\mathbb{Z}}_N$ representation ($SO(3)$ 
in our case) with periodic b.c., naturally transforming under 
the continuum Yang-Mills gauge group \cite{deForcrand:2002vs}. The adjoint 
partition function $Z({\beta}_A)$ should in fact be equivalent to $\tilde{Z}$ 
provided the $SO(3)$ native constraint 
$\,\sigma_c=\prod_{\overline{P}\in\partial c}\mathrm{sign}
(\mathrm{{Tr}_{F}}\,U_{\overline{P}})=1$ is satisfied for every elementary 
3-cube $c$, where $U_{\overline{P}}$ denotes the plaquettes belonging to its 
surface $\partial c$ 
\cite{Mack:1978rq,deForcrand:2002vs}:
${\mathbb{Z}}_2$ magnetic monopoles are suppressed and only closed 
${\mathbb{Z}}_2$ magnetic vortices winding around the boundaries are allowed,
i.e. in this limit adjoint actions dynamically allow all topological sectors 
which in the fundamental case must be fixed through twisted b.c.. 
Moreover, since the standard spontaneous center symmetry breaking argument 
for the 
deconfinement transition does not apply to the center-blind adjoint 
discretization the question whether $F$ behaves as an order parameter
is of major interest, also in light of recent studies for alternative 
descriptions of confinement in centerless theories \cite{Holland:2003jy}.

\begin{figure}[thb]
\begin{center}
\subfigure[]{\includegraphics[angle=0,width=0.49\textwidth]{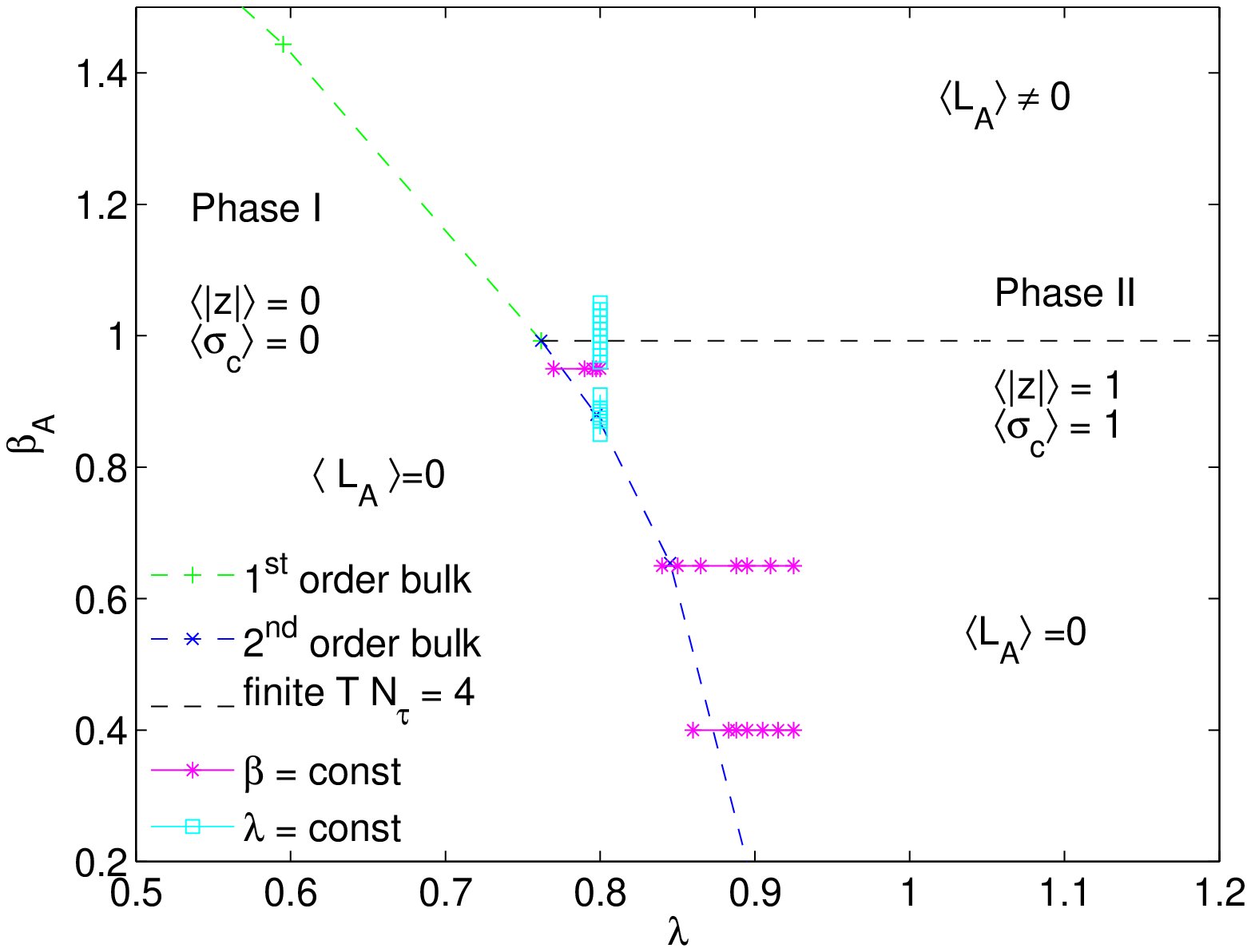}}
\subfigure[]{\includegraphics[angle=0,width=0.49\textwidth,%
height=0.36\textwidth]{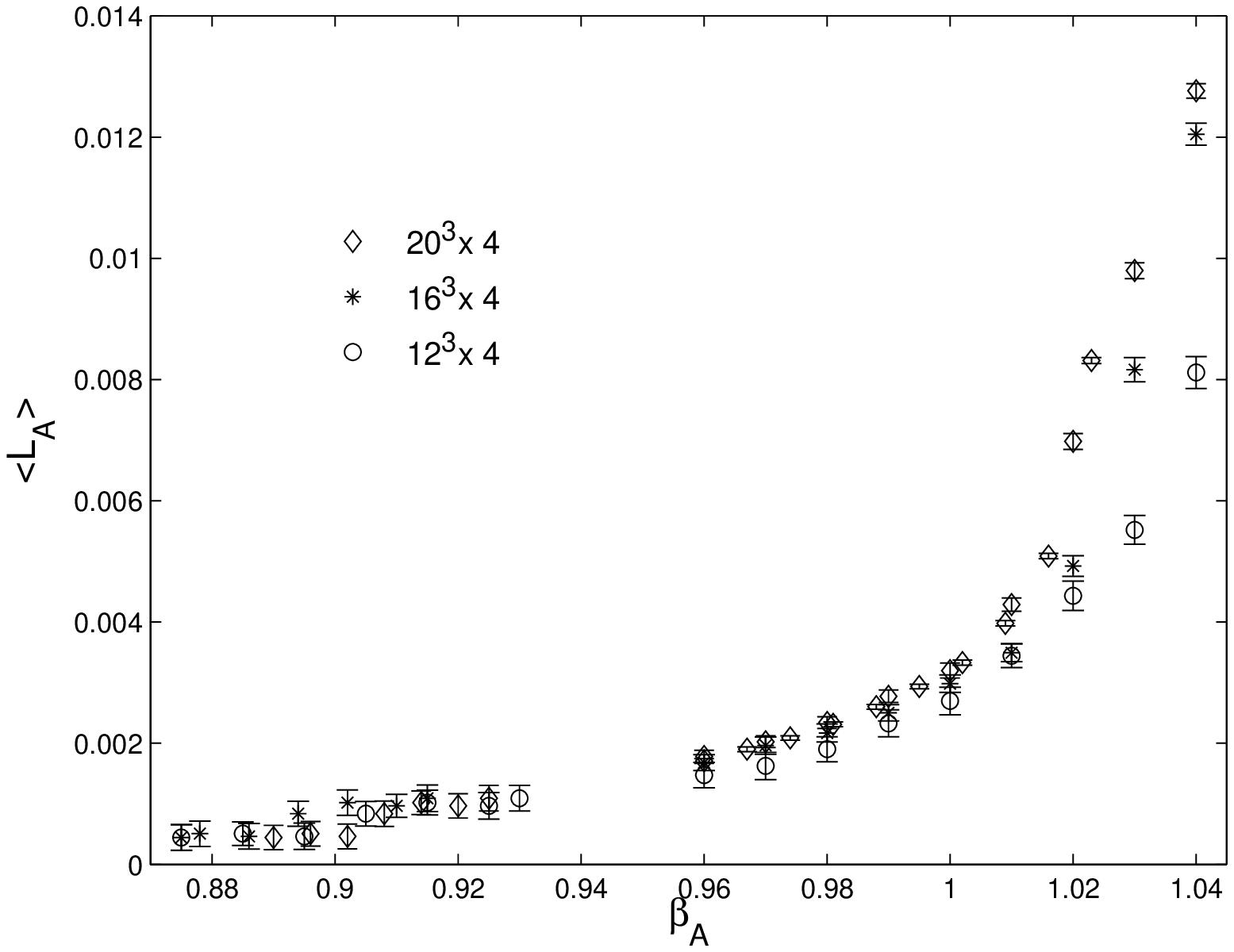}}
\end{center}
\caption{(a) Phase structure in the $\beta_A$-$\lambda$ plane 
with paths of couplings indicated as used in our simulations. 
(b) 
Adjoint Polyakov loop average
$\langle L_A \rangle$ versus $\beta_A$ at fixed $\lambda=0.8$ and 
$N_{\tau}=4$.} 
\label{phase}
\end{figure}

A practical obstacle one needs to overcome in investigating the adjoint
theory is the appearance of a bulk transition at some $\bar{\beta}_A$, 
separating a strong coupling chaotic phase (I) continuously connected 
with the fundamental action, where $\langle\sigma_c\rangle\simeq0$,
from a weak coupling ordered phase (II) where $\langle\sigma_c\rangle\simeq1$ 
\cite{Bhanot:1981eb,Greensite:1981hw,Halliday:1981te}. In phase II, where one 
wishes to exploit the relation between $Z({\beta}_A)$ and 
$\tilde{Z}$ mentioned above, high potential barriers separating
twist sectors suppress tunneling among them for local update
algorithms \cite{deForcrand:2002vs}. 
On the other hand a well-known result is that a center blind
$\mathbb{Z}_2$ monopole suppression term in the action 
$\lambda \sum_{c}(1-\sigma_{c})$ weakens the order of the bulk transition 
while moving it down into the strong coupling 
region (see the curved dashed line in
Fig.~\ref{phase}a) \cite{Halliday:1981te}.
For asymmetric lattice sizes $~N_{\tau} \times N_s^3, ~N_{\tau} \ll N_s~$
indications for a finite temperature critical line
$\beta^{crit}_A(\lambda, N_{\tau})$ within phase II
(horizontal dashed line in Fig.~\ref{phase}a 
for $N_{\tau}=4$)
have already been found from simulations at {\it fixed} twist 
\cite{Datta:1999np,Barresi:2003jq,Barresi:2004qa,Barresi:2006gq}.
%----------------------------------------------------------
%----------------------------------------------------------

In this paper we solve the problem of ergodic updates
through parallel tempering (PT) \cite{Hukushima:1995,Marinari:1996dh} 
for the adjoint Wilson action with $\mathbb{Z}_2$ monopole 
suppression
\begin{eqnarray} 
S=\beta_{A} \sum_{P} 
  \left(1-\frac{1}{3}\mathrm{Tr}_A U_{P}\right)
  +\lambda \sum_{c}(1-\sigma_{c})\,,
\label{ouraction}
\end{eqnarray}
where $U_P$ denotes the standard plaquette variable and 
$\mathrm{Tr}_A O= (\mathrm{Tr}_F O)^2-1 = \mathrm{Tr}_F (O^2)+1$ the adjoint 
trace. Our PT paths in the $\beta_A-\lambda$ plane
extend over the {\it second order} 
leg of the bulk transition straight into the 
inner region of phase II 
as shown in Fig.~\ref{phase}a, with two paths at fixed $\beta_A$, one
path with fixed $\lambda$, and one path consisting of a fixed-$\beta_A$
and a fixed-$\lambda$ part. In this way we avoid the 
necessity to cross high potential barriers.
This enables us to account for all twist sectors, restoring 
ergodicity, and to study $F$ at finite $T$. As a qualitative indicator 
of deconfinement we have also measured the adjoint 
Polyakov loop $\langle L_A \rangle = \langle{
{\sum_{\vec{x}}\mathrm{Tr}_{A}L(\vec{x})}}\rangle/(3 N_s^3)$, 
$L(\vec{x})= \prod_t U_4(\vec{x},t)$, which strictly speaking cannot 
behave as an order parameter for the deconfinement transition, 
delivering at most information on the screening length for the 
effective adjoint potential (see Fig.~\ref{phase}b). A preliminary report 
on the present investigation was presented in \cite{Burgio:2005xe};
further results on other observables 
as well as a detailed description of the implementation of the algorithm 
will appear in a separate publication \cite{Burgio:2006xj}. To our knowledge this is 
the first successful attempt to study the deconfinement transition of the 
$SO(3)$ lattice gauge theory via an ergodic simulation at large volume 
as well as the first determination of $F$ in the confined 
phase of a center-blind discretization of Yang-Mills theories.

In PT \cite{Hukushima:1995,Marinari:1996dh} we update $K$
configurations ${\cal{F}}_i$, $i = 1, \dots, K$, with couplings 
$(\lambda,\beta_A)_i$ swapping neighboring pairs %$i$, $i+1$
from smaller to larger values according to a Metropolis 
acceptance probability, satisfying detailed balance, $P_{\rm swap}(i,j) = 
\min[ 1, \mathrm{exp}(-\Delta S_{ij})]$, where 
\begin{equation}
\Delta S_{ij} = 
S[(\lambda,\beta_A)_i, {\cal{F}}_{i}] + S[(\lambda,\beta_A)_j, {\cal{F}}_{j}]
- S[(\lambda,\beta_A)_i, {\cal{F}}_{j}] - S[(\lambda,\beta_A)_j, 
{\cal{F}}_{i}]\,. 
\label{eq:Pswap-1}
\end{equation}
Compared to other methods \cite{Swendsen:1987ce,Berg:1991cf} the appeal of 
PT is its easy implementation both at criticality and away from it, needing 
no knowledge of re-weighting factors or other dynamical input. This is a
welcome property for us since we wish to go as far as possible 
from the bulk transition into phase II. In view of 
various experiences with simulated tempering one also expects PT to be more 
efficient than multi-canonical simulations. For the success of the method in 
the case under consideration the softening of the bulk transition to 
2$^{\rm nd}$ order is crucial, since we ``transport''  
tunneling from phase I at lower $\lambda$ into phase II at larger $\lambda$, 
where twist sectors are well defined but frozen. To work at low 
$\lambda$, i.e. through a 1$^{\rm st}$ order bulk transition, would make 
barriers too high and kill any hope of ergodicity at large volume, 
as experienced in \cite{deForcrand:2002vs} for $N_s > 8$. 

\begin{figure}[thb]
\begin{center}
\subfigure[]{\includegraphics[angle=0,width=0.49\textwidth,height=0.40\textwidth]%
{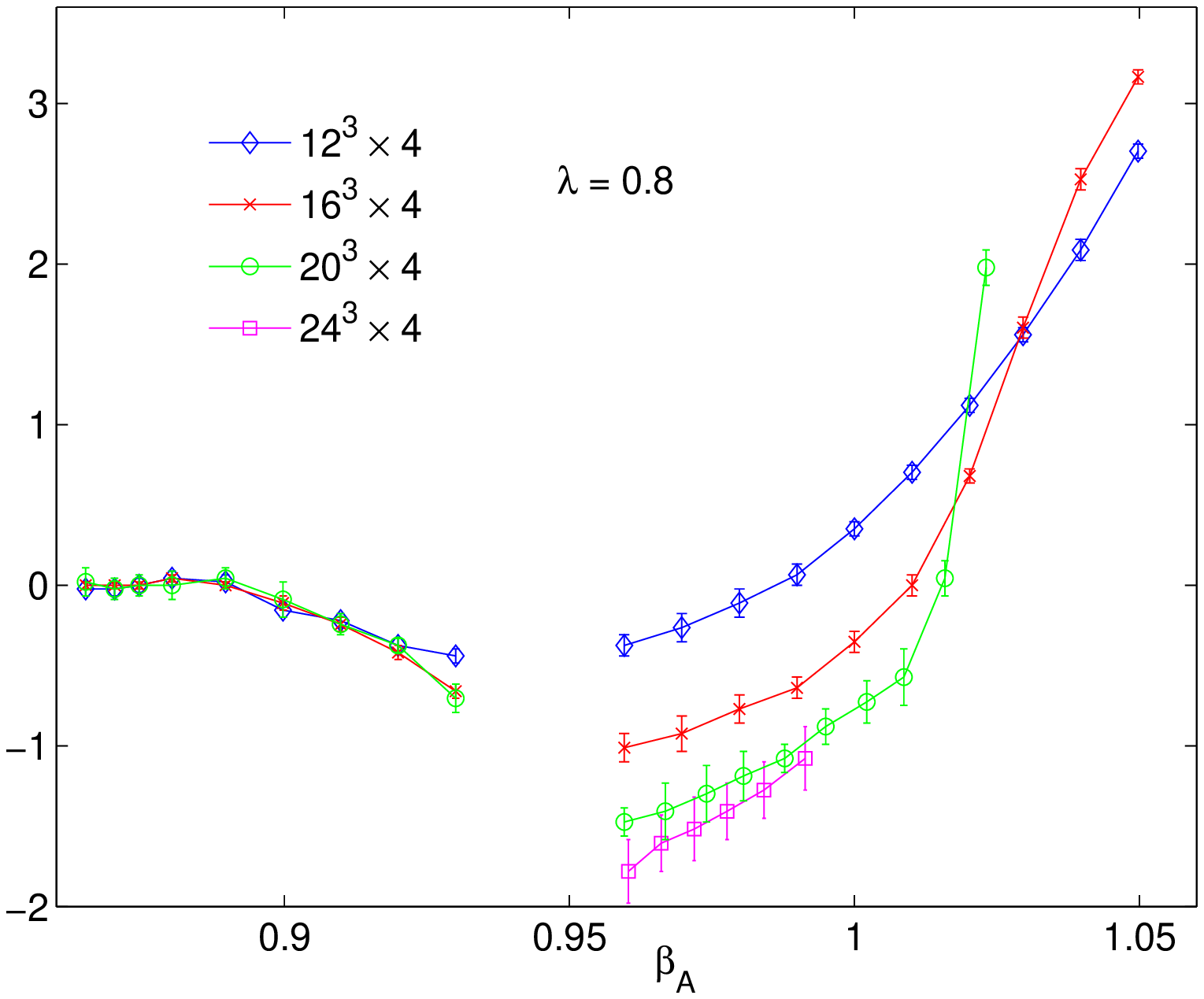}}
\subfigure[]{\includegraphics[angle=0,width=0.49\textwidth,height=0.40\textwidth]%
{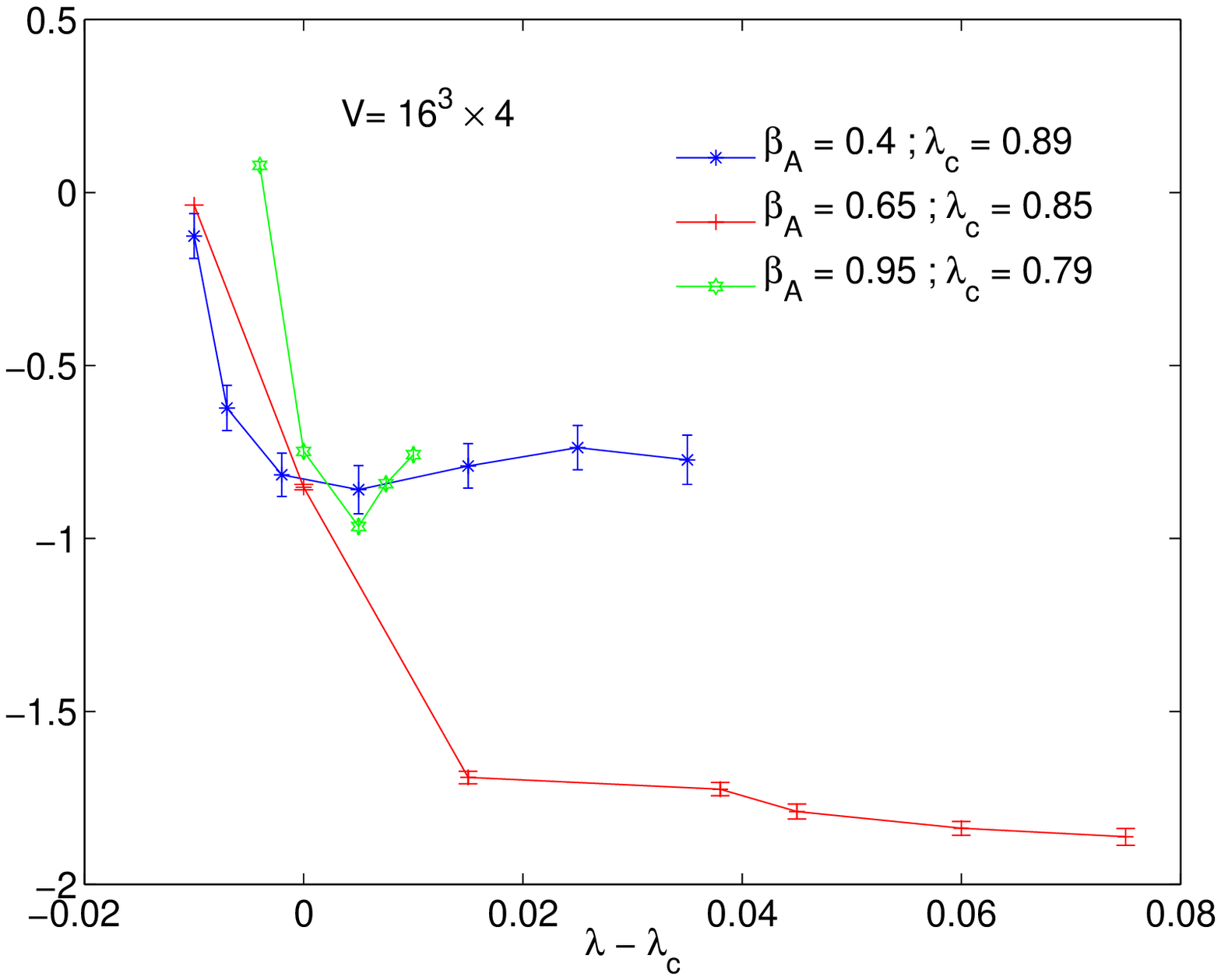}}
\end{center}
\caption{(a) $a N_\tau F$ versus $\beta_A$ at 
$\lambda=0.8$ and various $N_s$. (b) $a N_\tau F$ versus $\lambda-\lambda_c$
at fixed $\beta_A$ for $N_s=16$. $\lambda_c(\beta_A)$ denotes the position
of the bulk transition.}
\label{free1}
\end{figure}

Some care is of course necessary also with our method. 
In particular to maintain a sufficient swapping acceptance rate $\omega$ the 
distance between neighboring couplings must diminish with the volume. 
On the other hand to keep cross-correlations 
under control one does not wish the acceptance rate to be too high. We
have chosen to tune the parameters for each path and volume at hand
so to keep the acceptance rate roughly fixed at around $\omega = 12 \%$, 
a value for which we empirically find a good balance between auto- (in the
sense of freezing of the sectors) and cross-correlations.
For the paths in Fig~\ref{phase}a contributing to the left and right branches 
in Fig~\ref{free1}a details on the statistics, i.e. number of configurations 
$N$ and number of ensembles $K$, are given in Table~\ref{tab}. 
The paths at fixed $\beta_A = 0.4$ 
and 0.65 for $N_s = 16$ shown in Fig~\ref{free1}b were calculated with 
$K=7$ and $N=30000$ configurations. 
To remain on the safe side, in Fig~\ref{free1}a we have chosen not to quote 
the ensembles related to the end points of the paths since, having no further 
configurations to swap with, they might be affected by systematic errors. 
We wish however to stress that this has never been observed in the PT 
literature and we also have no indication that this might be the case.
Further details on the algorithm and a detailed analysis of correlations
will be reported in a forthcoming paper \cite{Burgio:2006xj}.
\begin{table}[thb]
\begin{center}
\subfigure[]{
\begin{tabular}{|c|c|c|}
\hline
$N_s$ & $K$& $N$\\
\hline
12 & 10& 30000\\
16 & 10& 30000\\
20 & 10& 30000\\
24 & 10& 30000\\
\hline
\end{tabular}}
\hspace*{1cm}
\subfigure[]{
\begin{tabular}{|c|c|c|}
\hline
$N_s$ & $K$& $N$\\
\hline
12 & 14 & 100000\\
16 & 14 & 100000\\
20 & 14 & 100000\\
24 & 10 & 100000\\
\hline
\end{tabular}}
\end{center}
\caption{Statistics achieved for the data given in Fig~\ref{free1}a.
The left branches correspond to (a) while the right branches together
with pieces at fixed $\beta_A=0.95$ and varying $\lambda$ correspond to
(b).}
\label{tab}
\end{table}
For the fixed $\lambda$ paths of Fig.~\ref{phase}a, along which main
simulations have been performed, we can fit very well
the step $\delta \beta_A$ needed to keep $\omega$ fixed with a 
law of the form
\begin{equation}
\delta \beta_{A}(\omega,N_s) \simeq \frac{\alpha(\omega)}{N_s^2}\,,
\end{equation}
where $\alpha(12\%)=2.15(3)$ in the $\beta_A = 0.95-1.09$ range considered, 
although we
expect it to change with the $(\lambda,\beta_A)$ window. Such scaling 
implies that in order to explore a fixed region $\Delta \beta_A$ of 
parameter space the number of ensembles will scale like
\begin{equation}
K \simeq \frac{\Delta \beta_A}{\alpha} N_s^2\,.
\label{scale}
\end{equation}
Therefore we cannot go too deep into phase II since the number of PT 
simulations will eventually become too large for the computational means
at our disposal.

The native $SO(3)$ temporal twists are given by $z_{i} \equiv N_s^{-2} 
\sum_{x_{j},\,x_{k}}\,\prod_{P\;\in\; \mathrm{plane}\; i,4}
\mathrm{sign Tr}_F U_P$, ($\epsilon_{ijk4}=1$) for $i=x,y,z$
\cite{deForcrand:2002vs,Barresi:2003jq}. Within phase II they are
well defined having values close to either 1 or -1 for each configuration. 
The partition functions restricted to fixed twist
are given as expectation values 
$\langle \nu_k\rangle \,=\,{Z|_{z=k}}/Z$ of the projectors 
\cite{'tHooft:1979uj}
\begin{eqnarray}
\nu_0  &=& {\frac{1}{8}}
\prod_{i=x,y,z}\lbrack1+\mathrm{sign}(z_{i})\rbrack\,,
\qquad 
\nu_1 ={\frac{1}{8}} \sum_{j={x,y,z}}\;
\prod_{i=x,y,z} \lbrack1+(1-2\delta_{i,j})\,\mathrm{sign}(z_{i})\rbrack\,,   
\nonumber\\
\nonumber\\
\nu_2 &=&{\frac{1}{8}}\sum_{j={x,y,z}}\;
\prod_{i=x,y,z} \lbrack1-(1-2\delta_{i,j})\,\mathrm{sign}(z_{i})\,,  
\qquad 
\nu_3 ={\frac{1}{8}}
\prod_{i=x,y,z}\lbrack1-\mathrm{sign}(z_{i})\rbrack\,.
\label{fractions}
\end{eqnarray}
The 2- and 3-twist sectors can of course 
only exist on $\mathbb{T}^3$ \cite{'tHooft:1979uj,deForcrand:2001nd}.
From Eq.~(\ref{fractions}), since for an adjoint theory a change of twist 
sector leaves the action unchanged, $\Delta U=0$ and:
\begin{equation}
F = -T \log{ {{Z_1}\over{3 Z_0}}}=
-{{1}\over{a N_\tau}} \log{ {{\langle\nu_1\rangle}
\over{3 \langle\nu_0\rangle}}}\,.
\label{eq6}
\end{equation}
The factor three in the denominator is again due to the 
three equivalent 1-twist sectors on $\mathbb{T}^3$ rather 
than one as on $\mathbb{R}^3$ \cite{deForcrand:2001nd,deForcrand:2002vs}. 
With such a choice $F$ will be 
zero if all twists are equally probable, i.e. on top of the bulk transition 
and everywhere in phase I, where twist sectors are however 
ill-defined due to the presence of open vortices. Eq.~(\ref{eq6}) 
obviously implies $F=0$ in the $T\to 0$ limit 
\cite{'tHooft:1979uj,Kovacs:2000sy,deForcrand:2000fi} as long as the
$\Delta S$ contribution remains bounded.
Fig.~\ref{free1}a shows numerical results for $F/T$ in lattice 
units at fixed $\lambda=0.8$ obtained 
along the two separated paths indicated in Fig.~\ref{phase}a. 
Errors are given combining statistical errors with 
auto and cross correlations.
%-----------------------------------------------------------------
%-----------------------------------------------------------------
The data start on top of the 2$^{\rm nd}$ order bulk 
transition and go upward to what we interpret as the finite $T$ 
deconfinement transition -- given the rapid growth of $\langle L_A \rangle$ 
(cf. Fig.~\ref{phase}b) as well as of $F$ and taking fixed-twist 
results for varying $N_\tau$ into account \cite{Barresi:2003jq,Barresi:2004qa,Barresi:2006gq}.

The behavior in phase I and just at the bulk transition as well as in the 
deconfined phase at large $\beta_A$ (upper phase II) is in agreement with the 
standard vortex arguments for confinement \cite{'tHooft:1979uj}: 
if vortices behave "chaotically" then $F$ should be zero and the theory 
confines, while as deconfinement occurs $F$ should rise as 
$F \sim \tilde{\sigma} N_s^2$, $\tilde{\sigma}(T)$ being the dual string 
tension. This is qualitatively in agreement with our data. While $F \approx 0$ 
close to the bulk transition, we find a strong rise of $F$ for 
$\beta_A \ge \beta_A^{crit} = 1.01(1)$ allowing to locate the 
finite-temperature transition in phase II.  As already explained above, 
we are not able to go too deep into the deconfined phase with the computing
facilities at our disposal. Thus, we cannot really check whether the data for 
$\beta_A \gg \beta_A^{crit}$ are consistent with the expected
$O(N_s^2)$ plateaus or whether they get saturated with respect to the 
thermodynamic limit, i.e. whether $\tilde{\sigma}$ can indeed be calculated.
The effort necessary to this purpose, assuming that the estimate in 
Eq.~(\ref{scale}) still works at higher $\beta_A$ and even taking into 
account that for higher volumes the asymptotic behavior should kick in 
earlier, we would need to simulate around 50 parallel ensembles for each 
volume, again for a statistics of at least $O(10^5)$ per configuration in 
each ensemble. For volumes with $N_s \geq 20$, for which finite size effects 
start to be reasonably small, this goes beyond the computational power at 
our disposal, although it should be manageable with a medium sized PC
cluster.
 
For $\beta_A \le \beta_A^{crit}$, i.e. throughout the confining region of
phase II, we find negative values for $F$ which stabilize at large 
3-volumes ($N_s \geq 20$). $F < 0$ comes as a surprise, meaning that vortex 
production is enhanced as compared to phase I. This is in contrast 
to what expected from arguments valid within the fundamental representation,
i.e. $F(T)=0$ throughout the confined phase. 
For an independent check we have carried out simulations at a few (fixed) 
lower 
$\beta_A$-values, varying $\lambda$ i.e. at somewhat lower temperatures
(see the horizontal paths drawn in Fig.~\ref{phase}a). 
The results are plotted in Fig.~\ref{free1}b. Again we
find $F<0$, but the plateau values do not behave monotonously 
as a function of $\beta_A$. Passing some minimum they increase again for 
decreasing $\beta_A$. This is compatible with the expectation that the free 
energy should go to zero in the zero-temperature limit 
\cite{'tHooft:1979uj,Kovacs:2000sy,deForcrand:2000fi}. A systematic 
extrapolation for different volumes and $N_\tau$ would be required to confirm 
this behavior and to decide whether $F/T$ itself vanishes or goes to a 
constant value. 

Let us draw the conclusions.
The main result of the present paper is the success in sampling the 
{\it full} partition function via ergodic PT Monte Carlo simulations 
and determining the free energy 
$F$ for the creation of a $\mathbb{Z}_2$ vortex in pure $SO(3)$ 
Yang-Mills theory at finite $T$. We have seen a clear indication
for a deconfinement transition consistent with earlier findings of a 
second order transition at fixed twist. Furthermore we  
find that $F$ does not vanish in the confined phase
at $T \neq 0$, vortex creation being enhanced throughout it; 
$F$ cannot therefore serve as an order parameter in a strict sense. 
This implies 
that the adjoint theory is unable to exhibit an order parameter 
for center symmetry breaking in any form, 
much like in the case of strictly centerless groups \cite{Holland:2003jy}. 
This is not in contradiction 
with the confining properties of the model, $F=0$ being a 
sufficient but not necessary condition for confinement away from the $T= 0$ 
limit \cite{'tHooft:1979uj,Tomboulis:1985}. Moreover while vortex 
suppression for $T<T_c$ would have been difficult to justify in light of 
the literature \cite{Greensite:2003bk} the 
vortex enhancement we observe does not contradict 
that they can play a r\^ole in describing confinement,
although one cannot speak of vortex condensation in the usual understanding.

Let us finish with a short remark on the universality problem we seem
to face. When identifying the partition functions  
$\tilde{Z}\simeq Z({\beta}_A)$, invoking universality for 
observables, one should be cautious. First of all, although 
${\mathbb{Z}}_2$ monopoles become suppressed in the continuum limit 
of $\tilde{Z}$, $\langle\sigma_c\rangle$ is still far from unity for 
the range of parameters commonly used in the simulations 
\cite{Kovacs:2000sy,deForcrand:2000fi,deForcrand:2001nd}
and open vortices might still dominate the partition function. 
Moreover $Z({\beta}_A)$ simply does not allow to define physical 
observables in the fundamental representation. Expectation
values of fundamental Wilson and Polyakov loops and all their correlators 
vanish identically for all $\beta_A$, i.e.
a fundamental string tension cannot be defined in a straightforward way.
To our knowledge bounds for $F$ from reflection positivity
or the connection with the electric flux have only been derived within the 
fundamental representation \cite{'tHooft:1979uj,Tomboulis:1985,
deForcrand:2001nd}. 
An interpretation of universality implying that any observable will 
assume the same value in any discretization is therefore trivially
contradicted by the above considerations. However, a slightly more 
conservative reading can agree with our findings: the truly physical 
properties measurable in ``experiments'' like glueball masses and the 
critical exponents at the transition should be reflected by physical 
observables which can be defined {\it irrespective} of the discretization 
chosen. 

We thank Ph. de Forcrand, J. Greensite, E.-M. 
Ilgenfritz, T. Kovacs, M. Pepe and U. J. Wiese for comments and discussions. 
G. B. acknowledges support from INFN.

\bibliography{bib.bib}
\end{document}